\documentclass[letterpaper,10pt,authoryear]{article}
\usepackage[]{graphicx}
\usepackage[toc,page]{appendix}
\usepackage{setspace} 

\usepackage{latexsym,amssymb,color,fullpage}
\usepackage{amsmath}
\usepackage{mathtools}
\usepackage{amsthm}

\RequirePackage{amsthm,amsmath,amsfonts,amssymb,arydshln}
\RequirePackage{graphicx}
\RequirePackage{threeparttable, tablefootnote}
\DeclareMathOperator{\tr}{tr}
\DeclareMathOperator*{\argmax}{argmax} 
\usepackage{bm} % to bold italic mathematical font

\theoremstyle{plain}

\usepackage{amsmath}
\usepackage{lmodern}
\usepackage{anyfontsize}
\usepackage{mathtools}
\usepackage{booktabs}
\usepackage{enumitem}

\newenvironment{keywords}[1][Key words:]{\begin{trivlist}

\item[\hskip \labelsep {\bfseries #1}]}{\end{trivlist}}
\newcommand{\T}{\mathsf{T}}

\usepackage{bm}

\usepackage{authblk}

\numberwithin{equation}{section}

\usepackage{algorithm}			  % I added
\usepackage[noend]{algpseudocode} % I added
\usepackage{algcompatible} 
\usepackage{arydshln}
\usepackage{booktabs}
\usepackage{enumitem}
\usepackage{graphicx}
\usepackage{float}
\usepackage{flafter} 
\usepackage[section]{placeins}
\usepackage{tabularx}

\newtheorem{thm}{Theorem}[subsection]% theorem counter resets every \subsection
\newcommand{\floor}[1]{\left\lfloor #1 \right\rfloor}

\usepackage[backend=biber, sorting=nyt, sortcites=true, style=apa, natbib=true, uniquename=false, bibstyle=apa]{biblatex}  %reference manager
\usepackage[pdfencoding=auto, bookmarks=true, hidelinks]{hyperref}
\usepackage{hyperref, cleveref}
\hypersetup{
     colorlinks   = true,
     citecolor    = blue,
     linkcolor    = blue
}

\title{A Network-Guided Penalized Regression with Application to Proteomics Data}
\author[1,2]{Seungjun Ahn}
\author[3]{Eun Jeong Oh}
\affil[1]{Department of Population Health Science and Policy, Icahn School of Medicine at Mount Sinai, New York, NY 10029, USA}
\affil[2]{Tisch Cancer Institute, Icahn School of Medicine at Mount Sinai, New York, NY 10029, USA}
\affil[3]{Institute of Health System Science, Feinstein Institutes for Medical Research, Manhasset, NY 11030, USA}
\date{}

\begin{document}

\maketitle

\allowdisplaybreaks

\begin{abstract}

Network theory has proven invaluable in unraveling complex protein interactions. Previous studies have employed statistical methods rooted in network theory, including the Gaussian graphical model, to infer networks among proteins, identifying hub proteins based on key structural properties of networks such as degree centrality. However, there has been limited research examining a prognostic role of hub proteins on outcomes, while adjusting for clinical covariates in the context of high-dimensional data. 
To address this gap, we propose a network-guided penalized regression method. 
First, we construct a network using the Gaussian graphical model to identify hub proteins. Next, we preserve these identified hub proteins along with clinically relevant factors, while applying adaptive Lasso to non-hub proteins for variable selection. 
Our network-guided estimators are shown to have variable selection consistency and asymptotic normality.
%However, identified hub proteins were not adjusted for any clinical covariates that may be associated with either the outcome variable or hub proteins themselves. Furthermore, in studies that have performed covariate-adjusted analysis by regressing identified hub nodes along with additional clinical covariates, variable selection techniques were not applied appropriately. To address these gaps, we propose a network-guided $\ell_1$-penalized regression method. First, we construct a network using a Gaussian graphical model to identify hub proteins. Next, we preserve hub proteins and clinically relevant factors while placing an adaptive Lasso penalty on non-hub proteins. 
Simulation results suggest that our method produces better results compared to existing methods and demonstrates promise for advancing biomarker identification in proteomics research. Lastly, we apply our method to the Clinical Proteomic Tumor Analysis Consortium (CPTAC) data and identified hub proteins that may serve as prognostic biomarkers for various diseases, including rare genetic disorders and immune checkpoint for cancer immunotherapy.

\end{abstract}

\begin{keywords}
Network analysis; Partial penalization; Adaptive Lasso; Proteomics data;
Protein interactions; Gaussian graphical model; Network connectivity; CPTAC.
\end{keywords}

\section{Introduction}

Proteins are key components of human cells and are involved in a diverse range of biological functions such as cell division and metabolism. Proteomics is the study of proteins on a large scale (i.e., proteome) and their interactions in a cell \citep{proteomics}. Recent advances in mass spectrometry (MS) technology has enabled simultaneous quantification of multiple protein expressions and identification of protein modification sites for proteomics research \citep{MS1, MS2}. The MS-based proteomics has been increasingly analyzed for biomarker discovery and disease monitoring in complex human diseases such as cancer \citep{proteomics.cancer1, proteomics.cancer2, CPTAC.petralia}, multiple sclerosis \citep{proteomics.MS}, Alzheimer's disease \citep{proteomics.AD}, and alcohol-related liver diseases \citep{proteomics.ALD}. More importantly, a proliferating number of proteomics studies has spurred development of statistical and bioinformatics methods to analyze the proteomics data. The majority of proteins do not act as independent entities. Instead, they work in concert (i.e., protein interactions) to induce and stabilize a range of cellular and physiological responses that include DNA replication, RNA transcription, protein translation, post-translational modification, targeted degradation, signal transduction, and cell cycle control \citep{PPI1}. 

The applications of network theory have proven instrumental in inferring the complex landscape of protein interactions through either correlation-based approaches or probabilistic graphical models, similar to other types of -omics disciplines \citep{GGM2}. Protein interactions can be represented as large interaction networks, wherein nodes symbolize proteins and edges denote pairwise interactions (co-expression), highlighting the presence of hub nodes based on network properties (e.g. proteins with higher degree centrality) \citep{PPI2}. Hub proteins play a pivotal role in maintaining the overall structure of a network. Thus, the removal of hub proteins may lead to a severe deterioration of network connectivity than that of non-hub proteins which has been referred to as centrality-lethality rule  \citep{hub1, hub2, hub4}. Furthermore, hub proteins are more likely to be encoded by genes associated with diseases than non-hub proteins \citep{hub3}.

Several authors have proposed various methods to estimate protein interactions, either through graphical model estimation or the Weighted Correlation Network Analysis (WGCNA) \citep{WGCNA}. Among them, \citet{glasso} introduced the graphical Lasso to estimate sparse undirected networks, which was subsequently validated using a small proteomics dataset. In related work, \citet{clinstudy3} performed a graphical Lasso-based network analysis on 36 protein biomarkers for imminent lung cancer diagnosis, demonstrating that the structural arrangement of a network changes according to the disease state (case or matched control) and identifying U-PAR as the central hub protein. On the other hand, \citet{clinstudy1} used the WGCNA to identify hub proteins, CD44 and PRDX1, within clusters of densely connected proteins (i.e., modules) which may serve as therapeutic targets for Alzheimer's disease. Recently, \citet{clinstudy2} fitted linear regressions to correlate continuous brain MRI outcomes and hub proteins identified from the WGCNA (``eigenproteins'' as described in the original paper) while accounting for additional clinical covariates such as age, sex, total/HDL cholesterol ratio, and prevalent cardiovascular diseases. 

However, the aforementioned studies have at least two of the following limitations: (1) identified hubs were not adjusted for clinical and demographic covariates; (2) the findings are primarily descriptive, and the investigation into the association between hubs and patient health outcomes remains unexplored, thereby limiting the interpretative framework of the study; and (3) a variable selection was not considered, resulting in a suboptimal prediction model characterized by an increased rate of false positives and reduced statistical power. Especially regarding the third limitation, a pertinent consideration arises – how should we address the retention of specific proteins variables and clinical covariates that may possess significant clinical and biological relevance, irrespective of their prior identification as hallmark biomarkers or genetic factors in existing studies? This brings into question the rationale for our \textit{``network-guided penalized regression}.''

Another major challenge in proteomics studies is the high-dimensionality of the covariate space. Recent developments in high-dimensional variable selection approaches include penalized regression methods, such as least absolute shrinkage and selection operator (Lasso) \citep{tibshirani}, adaptive Lasso \citep{zou2006}, smoothly clipped absolute deviation method \citep{fan}, elastic net \citep{enet}, nonnegative garrote \citep{yuan}, and many others.  
Recent studies \citep{villanueva, xu2024genetic} have regressed all proteins in the same prediction model and derived a model with a reduced number of proteins using penalization techniques, including Lasso and elastic net. However, this approach does not consider that a set of proteins interacts with each other as a network. Furthermore, penalizing all proteins is not appropriate when certain variables, such as hubs proteins (for preserving the overall network structure) and clinical covariates (for their clinical importance and potential confounding), should remain in the model. Another study by \citet{tutz2009penalized} proposed utilizing the correlation between predictors explicitly in the penalty term. However, their method relies on marginal correlations, which may fail to capture the conditional dependencies among variables that underlie network structures. In contrast, our approach uses partial correlations to reflect direct associations while adjusting for the effects of other protein variables, providing a more biologically meaningful representation of molecular networks.

%Recently, \cite{klau} attempted to prioritize blocks of variables in omics data based on the sequence determined by clinicians’ preferences or knowledge. Despite their efforts to incorporate clinical priorities, this approach would become labor-intensive unless pre-defined blocks already exist, especially when dealing with hundreds or thousands of variables.

In the present work, we propose to incorporate network knowledge into variable selection with adaptive Lasso. Specifically, our proposed network-guided penalization procedure retains hub proteins and clinical covariates, while applying an adaptive Lasso penalty to non-hub proteins. %The overarching goals of this study are to: (1) characterize a network with sparse Gaussian graphical model to differentiate hub and non-hub proteins; and (2) preserve important hub nodes identified through network estimation and essential clinical variables in a regression framework, while selectively allowing non-hub proteins to remain in the model based on their predictive capabilities. 
The overarching objective of this study is to introduce a method that differentiates hubs from non-hubs and is designed to assess the covariate-adjusted effect of hubs on patient health outcomes with the removal of irrelevant non-hubs. 
This dual strategy will help preserve the overall network structure by retaining hubs and also enhance the model predictive accuracy by properly adjusting for clinical confounders and penalizing non-hubs for variable selection. 
%The goal of our method is to explain the variability in the outcome using important hub proteins and potential confounders along with the selected non-hub proteins based on data-driven variable selection. 

This article is organized into five main sections. Section 1 provides background and motivations. Section 2 covers network estimation, network-guided penalization, and the asymptotic behaviors of the proposed estimators. In Section 3, we present performance metrics from simulation experiments, comparing our method with existing alternatives. In Section 4, we apply our proposed method to proteomics data from the National Cancer Institute (NCI) Clinical Proteomic Tumor Analysis Consortium (CPTAC). Finally, we wrap up our discussion by addressing challenges, limitations, and future directions in Section 5.

\section{Methods}

Consider a finite population of $n$ subjects. %and let $i = 1, \dots , n$ denote each subject in the study. 
Let $Y$ be an outcome of interest, $\bm{X} = \{X_{1}, \dots , X_{p}\}$ be a vector of proteins, $\bm{Z} = \{Z_{1}, \dots , Z_{c}\}$ be a vector of potential confounders that need to be adjusted in the regression, such as age, gender, and other related diseases and conditions at baseline. In our data example, $\bm{X}$ is high-dimensional, whereas the dimension of $\bm{Z}$ is low or moderate. 

\subsection{Network Estimation With Sparse Gaussian Graphical Model}

Our idea is built upon the Gaussian graphical model (GGM) \citep{GGM1} to estimate a PPI network, where an edge represents conditional dependency of a pair of nodes (proteins) after controlling for all other nodes in a network. In a GGM network, the weight of an edge is the partial correlation and represent whether or not and how strongly the two nodes co-occur. Thus, the network structure is decided by an estimation of partial correlations. 

We assume $\bm{X}$ follows a multivariate normal distribution:
\begin{align*} 
\bm{X} \sim N_p(\bm \mu, \bm \Sigma),
\end{align*}
where $\bm{\mu} = (\mu_{1}, \dots, \mu_{p})$ is a vector of means for each protein and 
%$\bm\Sigma = \{\sigma_{jk}\}$, $1 \leq j \leq k \leq p$
$\bm\Sigma = [\Sigma_{jk}]_{1\leq j,k \leq p}$
is a $p \times p$ variance-covariance matrix that is positive definite. The inverse of $\bm\Sigma$ is a precision matrix (or concentration matrix), denoted by $\bm\Sigma^{-1} = \bm\Theta =[\theta_{jk}]_{1\leq j,k \leq p}$. The off-diagonal elements of precision matrix can be standardized with a sign reversal to calculate the partial correlation of two proteins $X_{j}$ and $X_{k}$, conditional on all other proteins in $\bm{X}$ \citep{GGM2}, which is our focus in this section,
\begin{align}
    \rho_{X_{j},X_{k} | \bm{X}_{-j, -k}} = \frac{-\theta_{jk}}{\sqrt{\theta_{jj}\theta_{kk}}},
    \label{netreg:pcm}
\end{align}
where $\bm{X}_{-j, -k}$ is a set of proteins without $j$ and $k$, and $\theta_{jk}$ denotes the corresponding element of $\bm\Theta$. Furthermore, $\bm\Theta$ can be constructed as a network with protein nodes that are connected by edges when $\rho_{X_{j},X_{k} | \bm{X}_{-j, -k}} \neq 0$. 

As in most ``-omics'' disciplines, there are generally hundreds of samples, while each sample has thousands of proteins \citep{GGM.highdim}. In the high-dimensional setting where $n \ll p$, a maximum likelihood estimation of $\bm\Sigma$ may not be accurate due to singularity (i.e., $\det(\bm\Sigma) = 0)$ \citep{GGM.sparse1}. Thereby, a range of regularization methods have been proposed to estimate a sparse GGM, which bypass the issue of $n \ll p$ and non-invertible $\bm\Sigma$ \citep{GGM.type1, GGM.type2, glasso}.

\begin{figure}[ht]
\centering
\includegraphics[scale=0.3]{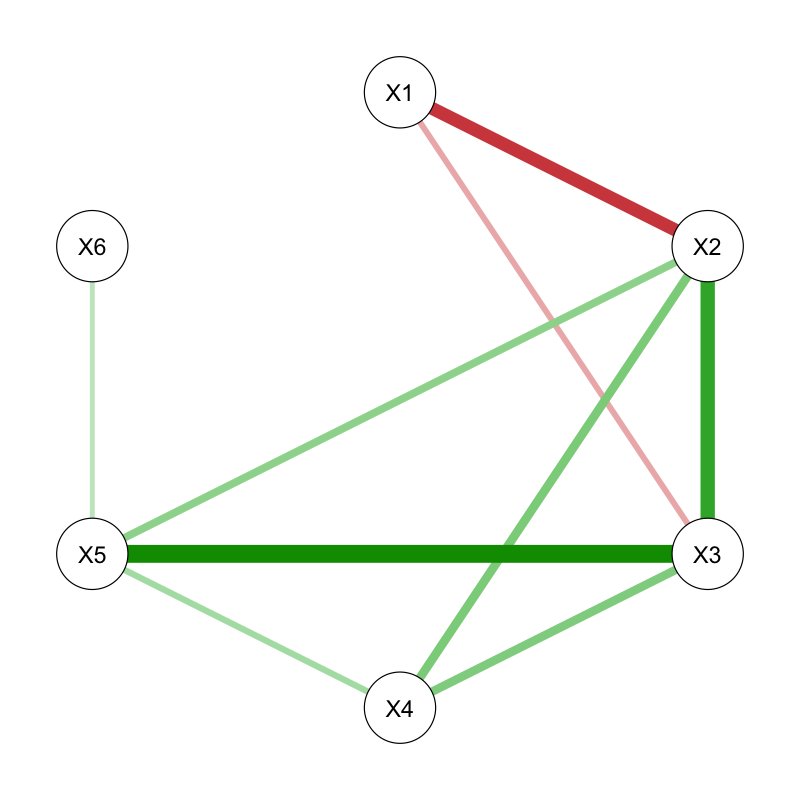}
\caption{An example network plot to visualize a network with $p = 6$ proteins based on partial correlation estimates from a graphical Lasso algorithm in combination with an extended Bayesian information criterion. Any missing edge between nodes (e.g. $X_{1}-X_{4}$, $X_{1}-X_{5}$, and $X_{1}-X_{6}$) corresponds to a partial correlation estimates of exactly zero in $\hat{\bm\Theta}$.}
\label{fig1}
\end{figure}

In the present paper, we consider the graphical Lasso (GL) \citep{glasso} to impose a sparsity on GGM by penalizing $\ell_{1}$-norm of the elements of $\bm\Theta$. The GL estimates $\bm\Theta$ by maximizing the following penalized log-likelihood:

\begin{align} 
\hat{\bm\Theta} = \argmax_{\bm\Theta} \bigl\{ \log \det(\bm\Theta) - \tr(\hat{\bm\Sigma}\bm\Theta) - \lambda \lVert \bm\Theta \rVert_{1} \bigr\},
\label{netreg:GL}
\end{align}
where $\hat{\bm\Sigma}$ is an empirical variance-covariance matrix, $ \lVert \bm\Theta \rVert_{1} = \sum_{j \neq k} |\theta_{j,k}| $ denotes the sum of absolute value of edges, and $\lambda$ denotes a tuning parameter. The optimal $\lambda$ can be chosen based on the extended Bayesian information criterion (eBIC) \citep{ebic}.
The eBIC is expressed as
\begin{align*} 
\text{eBIC} = -2 \ell(\bm\Theta) + E \log(n) + 4\gamma E \log(p),
\end{align*}
where $\ell(\bm\Theta)$ is a penalized log-likelihood to estimate $\hat{\bm\Theta}$ in equation \eqref{netreg:GL}, $E$ is a number of edges or non-zero elements of $\bm\Theta$, and $\gamma \in [0,1]$ denotes a non-negative eBIC hyperparameter. Of note, the eBIC becomes an ordinary BIC when $\gamma = 0$ and a higher value of $\gamma$ leads to a greater sparseness by removing more edges \citep{GGM2}. Following equation \eqref{netreg:pcm}, $\hat{\rho}_{jk}$'s are calculated based on the elements of resulted $\hat{\bm\Theta}$. See Figure \ref{fig1} for an example of the network constructed as described.

The network centrality has been studied to measure the extent of biological or topological importance that a node has in a network \citep{centrality1, centrality2}. For each protein $k$, the network centrality (degree centrality in continuous scale; $\hat{\phi}_{k}$) is calculated as the marginal sum of the association matrix.
\begin{equation*}
    \hat{\phi}_{k} = \sum_{j=1}^{p} | \hat{\rho}_{jk} | ,
\end{equation*}
where $k = 1, \dots , p$. We define protein nodes with higher $\hat{\phi}_{k}$ relative to others as hub proteins, where the number of hub proteins $h < p$ depends on user-specified parameters, $\delta$ and $\tau$, by taking $h = \min(\floor{p \delta}, \tau)$. More details are discussed in the following subsection.

\subsection{Network-Guided $\ell_1$-Penalization}
\label{sc:network}

Suppose we observe data from $n$ individuals. For each individual, the data is of the form $\{ \bm Z, \bm X, Y \}$. When the goal is to regress $\bm Z, \bm X$ on $Y$, the following model is usually considered:
\begin{align}
%Y = \Phi(\bm Z, \bm X) \eta + \epsilon,
Y = \mu + \bm Z \bm \zeta + \bm X \bm \eta + \epsilon,
\label{genmodel}
\end{align}
where $\mu$ is the intercept, $\bm \zeta$ and $\bm \eta$ are the coefficients for $\bm Z$ and $\bm X$, respectively, and $\epsilon$ is the error component assumed to be normally distributed around zero with constant variance $\sigma^2$.

In this project, we decompose $\bm X$ into two parts: hub proteins, denoted by $\bm H \in \mathbb{R}^{h}$, and non-hub proteins, denoted by $\bm N = \bm X \setminus \bm H \in \mathbb{R}^{q}$, such that $\bm\eta = (\bm\eta_1, \bm\eta_2)$, where $\bm\eta_1$ and  $\bm\eta_2$ are the coefficients of $\bm H$ and $\bm N$, respectively.
The network-guided $\ell_1$-penalization procedure aims to adjust the level of penalization on non-hub proteins $\bm N \in \mathbb{R}^q$, while preserving hub proteins and and potential confounders along with unpenalized intercept, denoted by $\bm U = (1, \bm Z, \bm H) \in \mathbb{R}^{t}$, where $t=h+c+1$, in the model to be adjusted for. Thus, model \eqref{genmodel} can be re-written as
\begin{align}
    Y 
    & = \mu + \bm Z \bm\zeta + (\bm H, \bm N) (\bm\eta_1, \bm\eta_2)^\T + \epsilon \nonumber \\
    & = (1, \bm Z, \bm H) (\mu, \bm\zeta, \bm\eta_1)^\T + \bm N \bm\eta_2 + \epsilon \nonumber \\
    & = \bm U \bm \alpha + \bm N \bm \beta + \epsilon, \label{fin.model}
\end{align}
where $\bm \alpha = (\mu, \bm\zeta, \bm\eta_1)$ and $\bm \beta = \bm\eta_2$ are the corresponding coefficients for $\bm U$ and $\bm N$, respectively.

%Due to the high-dimensionality of proteomics data, the conventional least-squares method is no longer feasible.
To deal with high-dimensional data, we propose a regression approach with a Lasso-type penalty. 
The network-guided $\ell_1$-penalization estimates $(\bm{\hat{\alpha}}_n, \bm{\hat{\beta}}_n)$ are obtained by minimizing the following objective function:
\[
L_n (\bm\alpha, \bm\beta) =  \| Y -  \bm U \bm \alpha - 
\bm N \bm \beta \|_2^2 + \lambda_n \sum_{j=1}^{q} w_j |\beta_j|,
\]
where $\lambda_n$ is a non-negative tuning parameter that controls model complexity and $w_j \geq 0$ is the weight for adjusting the level of penalization on $\beta_{j}$.
%When all $w_j$'s are set to 0, it becomes regular linear regression, while when all $w_j$'s are set to 1, it becomes a partial Lasso regression. 
In this project, we apply adaptive Lasso \citep{zou2006} to shrink coefficients of non-hub proteins such that only significant ones remain in the model, while keeping hub proteins and clinical covariates. %\textcolor{red}{EJ: For adaptive Lasso, the weights are not excessively big for nonzero coefficients, and the weights are not too small for zero coefficients, by setting them to be inversely proportional to the magnitude of a root-$n$-consistent estimator, $\tilde{\bm\beta}_n$.}  This can be accomplished by establishing the weight vector as $\bm{\hat w} = |\tilde{\bm\beta}_n|^{-\nu}$ for some $\nu > 0$.
In adaptive Lasso, the weight vector is defined as $\bm{\hat w} = |\tilde{\bm\beta}_n|^{-\nu}$ for some $\nu > 0$, where $\tilde{\bm\beta}_n$ is any root-$n$-consistent estimator. This imposes heavier penalties on covariates with smaller coefficients.
In practice, we use perturbed elastic net estimates for $\tilde{\bm\beta}_n$, following \citet{zouzhang2009}. The 5- or 10-fold cross-validation can be used to select an optimal pair of $(\nu, \lambda_n)$. \\

The formula $h = \min(\floor{p \delta}, \tau)$ which is used to identify $\bm H$ helps control that the number of hub proteins depends on a user-specified proportion to the size of $\bm X$ with the pre-specified positive constant $\tau$. In this study, we set $\tau = \floor{(p+20)/16} $, such that the dimension of non-penalized terms is moderate. It is essential that starting with the minimal size of $\bm H$ is desired due to the nature of a partial penalization. Even if all essential proteins were not classified as $\bm H$, they would still undergo evaluation as $\bm N$ through a penalization method and could remain in the final model if they are shown to be predictive of outcomes in a data-driven manner.

We assume the following two regularity conditions:
\begin{enumerate}
	\item[(A1)]  $\epsilon \overset{\Delta}{=} Y - \bm{U}\bm\alpha_0 - \bm{N}\bm\beta_0$ has mean zero and finite variance $\sigma^2$, and is independent of $(\bm U, \bm N)$.
\label{reg.condition1}
	\item[(A2)] $n^{-1}(\bm U, \bm N)^\T (\bm U, \bm N) \rightarrow \bm C$, where $\bm C$ is positive definite. \label{reg.condition2}
\end{enumerate}

Let $\mathcal{J} = \left \{ j: \beta_{0j} \neq 0, j=1,\ldots, q \right \}$  be the true active set of variables in $\bm N$, and assume that $|\mathcal{J}| = r < q$. Denote the estimated active set of variables by $\hat{\mathcal{J}}_n = \left \{ j: \hat{\beta}_{nj} \neq 0, j=1, \ldots, q \right \}$. Let 
${\bm{\beta}}_{0\mathcal{J}} = \{{\beta}_{0j}: j\in\mathcal{J}\}$ and  $\hat{\bm{\beta}}_{n\mathcal{J}} = \{\hat{\beta}_{nj}: j\in\mathcal{J}\}$. 
Denote $\bm{\theta}=(\bm\alpha^\T,\bm\beta^\T)^\T$ for any $\bm\alpha\in\mathbb{R}^{t}$, $\bm\beta\in\mathbb{R}^{q}$. Then $\mathcal{S} = \left \{1, 2, \ldots, t \right \} \cup \left \{s: \theta_{0s} \neq 0, \,s = t+1, \ldots, t+q \right \}$ is the true active set of variables in $(\bm U, \bm N)$, and thus $\mathcal{J}$ is always the subset of $\mathcal{S}$. 
Denote $\bm{C}_{\mathcal{S}} \in \mathbb{R}^{(t+r) \times (t+r)}$ is the top-left block matrix (i.e., sub-matrix) of $\bm C \in \mathbb{R}^{(t+q) \times (t+q)}$. In the following, we demonstrate the oracle property of our estimators.

\begin{thm}
Suppose $\lambda_n = o(\sqrt{n})$ and $\lambda_n n^{(\nu-1)/2} \rightarrow \infty$. Then under model \eqref{fin.model} and regularity conditions (A1)--(A2), the network-guided adaptive Lasso estimators satisfy the following properties:

\begin{itemize}
\item[i)] (variable selection consistency) $\lim_n P( \hat{\mathcal{J}}_n = \mathcal{J} ) = 1$,
\item[ii)] (joint asymptotic normality) 
\[
\sqrt{n} 
\begin{pmatrix}
\hat{\bm{\alpha}}_n - \bm{\alpha}_0\\ 
\hat{\bm{\beta}}_{n\mathcal{J}} - \bm{\beta}_{0\mathcal{J}}
\end{pmatrix} \rightarrow_d N(\bm 0, \sigma^2 \bm C_\mathcal{S}^{-1} ).
\]
\end{itemize}
\end{thm}

The theorem above implies that the network-guided adaptive Lasso estimators enjoys variable selection consistency and asymptotic normality. The proof is deferred to the Appendix.

\section{Simulation Experiments}

In this section, simulation studies are conducted to compare the proposed network-guided (NG) adaptive Lasso estimators with other existing alternatives and evaluate model performance using various metrics. We make a comparison with the adaptive Lasso (aLasso), Lasso, elastic net (enet), and ridge regression models, as well as the correlation-based penalized estimators (CBPE) proposed by \citet{tutz2009penalized}. For each method, the 5-fold cross-validation was used to select the optimal tuning parameters.

\subsection{Settings}

We generated $\bm X$ from the multivariate normal distribution $N_p(\bm 0, \bm \Sigma)$ with the correlation structure 
$\bm \Sigma = [\Sigma_{jk}]_{1\leq j,k \leq p}$, where $\Sigma_{jk}$ is 1 if $j=k$, 0.9 if $j \in \{1,2,3,4\} \neq k$, and $0.9^{|j-k|}$ if $j \in \{5, \ldots, p\} \neq k$.  %to make the first four proteins to serve as hub proteins.
Three potential confounders were generated as follows: $Z_1 \sim U(0,1)$, $Z_2 \sim \textup{Bernoulli}(0.25)$, and $Z_3 \sim \textup{Bernoulli}(0.65)$. The outcome variable was generated according to the model \eqref{genmodel} with $\sigma=1$, $\mu = 0.5$, $\bm \zeta = (2.5, 2.5, 2.5)$ along with the two different scenarios for $\bm \eta$:
\begin{itemize}
\item[1.] Strong signal: $\bm\eta = (3.5_5, 0_5, -1.5_5, 0_{p-15})$
\item[2.] Weak signal: $\bm\eta = (1, -0.8, 0.6, 0, 0, -1.5, -0.5, 1.2, 0_{p-8})$
\end{itemize}

Different combinations of sample size and dimension (network size), denoted as $(n, p)$ = $(50, 60)$, $(100, 60)$, $(100, 300)$, were considered. Following the terms used in \citet{monti2021sparse}, each sample size/dimension combination represents moderate-high-dimensional ($n<p$), low-high-dimensional ($n>p$), and high-dimensional setting ($n \ll p$), respectively. The high-dimensional setting is often observed in proteomics studies, as in most ``-omics'' disciplines. For each setting, we repeated the simulation $100$ times.

\subsection{Performance metrics}

The predictive model performance is mainly evaluated using the root-mean-squared error (RMSE) and calibration slope (CSL). Overall variable selection performance was assessed by the F1 score, defined as
\[
\text{F1 score} =\frac{2 \cdot \text{TP}}{ 2 \cdot \text{TP} + \text{FP} +\text{FN}},
\]
and the Matthews correlation coefficient (MCC) proposed by \citet{mcc}, defined as
\begin{align*}
    \text{MCC} = \frac{\text{TP} \cdot \text{TN} - \text{FP} \cdot \text{FN} }{\sqrt{(\text{TP}+\text{FP})\cdot(\text{TP}+\text{FN})\cdot(\text{TN}+\text{FP})\cdot(\text{TN}+\text{FN})}},
\end{align*}
where TP, TN, FP, and FN are true negatives, true negatives, false negatives, and false positives, respectively. 
The performance measures were evaluated on an independent test set of size $1,000$.

\subsection{Simulation results}

Tables \ref{sim.res1}--\ref{sim.res2} present model performance metrics under various settings based on 100 simulation replicates. For each metric, the mean value is reported along with the standard deviation in parentheses. In the strong signal case (Table \ref{sim.res1}), with different specifications of $\delta$, our proposed NG method consistently outperformed other existing methods, in terms of lower RMSE, better calibration, higher F1 score, and higher MCC. Especially when $n> p$, aLasso and our NG method performed well, in terms of calibration slope nearly close to an ideal value of 1 and a very high MCC. 

However, it is worth noting that in settings where the number of proteins is greater or significantly greater than the number of observations (i.e., $n < p$ or $n \ll p$), our NG method showed much better performances than the alternative methods. For example, when $n<p$, the NG method had a F1 score of 0.81, which was higher than the rest of the methods, ranging from 0.34 (ridge or CBPE) to 0.71 (aLasso). 
Furthermore, the proposed NG method demonstrated a smaller standard deviation of RMSE, compared to the competing methods. When $n \ll p$, the RMSE standard deviations were 0.11 (NG, $\delta=0.01$) and 0.13 (NG, $\delta=0.02$ or $\delta=0.03$), which were smaller than 0.90 (ridge) or 0.46 (aLasso). Additionally, ridge, enet, and CBPE showed poor performance, characterized by one or more of the following: high RMSE, calibration slope far from 1, low F1 score, or low MCC.

Similarly, in the weak signal case (Table \ref{sim.res2}), our proposed NG method continued to outperform the other methods. For instance, in the high-dimensional setting $n\ll p$, the RMSE of the NG method with $\delta=0.02$ was 0.19, which was less than half of that of aLasso (RMSE, 0.45), Lasso (RMSE, 0.46), and enet (RMSE, 0.53), and almost one-tenth of ridge (RMSE, 2.14) or CBPE (RMSE, 2.48). In all settings, the NG method showed lower RMSE, better calibration, higher F1 score, and higher MCC. 

In all cases, our method consistently performed well even with a small $\delta$. As discussed in Section \ref{sc:network}, proteins that may have been initially missed out based our network estimation can still be included in the final model through variable selection if they show predictive potential, thereby demonstrating good overall model performance.

\begin{table*}
\centering
\caption{Simulation results under strong signal case using network-guided (NG) method, adaptive Lasso (aLasso), Lasso, elastic net (enet), ridge regression, and correlation-based penalized estimators (CBPE). The best results are highlighted in boldface.}
\label{sim.res1}
\begin{tabular}{@{}l cc l cccc @{}}
\toprule
Setting & $n$ & $p$ & Method & RMSE & CSL & F1 score & MCC \\
\midrule 
$n<p$ & 50 & 60 & NG ($\delta=0.06$)  & 1.33 (0.59) & \textbf{1.01} (0.01) & \textbf{0.81} (0.11) & 0.75 (0.15) \\
   &    && NG ($\delta=0.08$)  & \textbf{1.32} (0.57) & \textbf{1.01} (0.01) & \textbf{0.81} (0.11) & \textbf{0.77} (0.14) \\
   &    && NG ($\delta=0.10$)  & \textbf{1.32} (0.63) & \textbf{1.01} (0.01) & \textbf{0.81} (0.10) & 0.76 (0.13) \\
   &    && aLasso              & 2.89 (0.78) & 1.04 (0.03) & 0.71 (0.15) & 0.66 (0.19) \\
   &    && Lasso               & 1.88 (0.71) & 1.03 (0.02) & 0.64 (0.10) & 0.56 (0.14) \\
   &    && enet                & 2.22 (0.58) & 1.04 (0.02) & 0.52 (0.07) & 0.40 (0.11) \\
   &    && ridge               & 8.16 (0.72) & 1.63 (0.14) & 0.34 (0.00) & -- \\
   &    && CBPE                & 2.77 (0.31) & 1.04 (0.03) & 0.34 (0.00) & -- \\
\midrule
$n > p$ & 100 & 60 & NG ($\delta=0.06$)  & \textbf{0.67} (0.10) & \textbf{1.01} (0.00) & \textbf{0.99} (0.03) & \textbf{0.98} (0.03) \\
   &    && NG ($\delta=0.08$)  & 0.68 (0.11) & \textbf{1.01} (0.00) & 0.98 (0.03) & \textbf{0.98} (0.03) \\
   &    && NG ($\delta=0.10$)  & 0.70 (0.11) & \textbf{1.01} (0.00) & 0.95 (0.03) & 0.94 (0.04) \\
   &    && aLasso              & 0.70 (0.10) & \textbf{1.01} (0.00) & 0.98 (0.04) & \textbf{0.98} (0.04) \\
   &    && Lasso               & 0.74 (0.12) & 1.02 (0.00) & 0.71 (0.07) & 0.65 (0.08) \\
   &    && enet                & 0.89 (0.14) & 1.02 (0.00) & 0.51 (0.05) & 0.41 (0.07) \\
   &    && ridge               & 0.94 (0.11) & 1.02 (0.01) & 0.34 (0.00) & -- \\
   &    && CBPE                & 1.57 (0.17) & 1.02 (0.01) & 0.34 (0.00) & -- \\
\midrule
$n \ll p$ &100 & 300 & NG ($\delta=0.01$)  & \textbf{1.08} (0.11) & \textbf{1.00} (0.00) & \textbf{0.97} (0.03) & \textbf{0.97} (0.03) \\
   &    && NG ($\delta=0.02$)  & 1.15 (0.13) & \textbf{1.00} (0.00) & 0.88 (0.03) & 0.87 (0.03) \\
   &    && NG ($\delta=0.03$)  & 1.19 (0.13) & \textbf{1.00} (0.00) & 0.79 (0.03) & 0.78 (0.03) \\
   &    && aLasso              & 2.42 (0.46) & 1.02 (0.01) & 0.85 (0.06) & 0.86 (0.05) \\
   &    && Lasso               & 1.15 (0.21) & 1.02 (0.00) & 0.92 (0.06) & 0.92 (0.05) \\
   &    && enet                & 1.22 (0.23) & 1.02 (0.00) & 0.85 (0.07) & 0.85 (0.07) \\
   &    && ridge               & 9.87 (0.90) & 1.45 (0.07) & 0.08 (0.00) & -- \\
   &    && CBPE                & 5.29 (0.33) & 1.07 (0.02) & 0.08 (0.00) & -- \\
\bottomrule
\end{tabular}
\begin{tablenotes}\scriptsize
\item [] Abbreviations: RMSE, root-mean-squared error; CSL, calibration slope
\item [] $\delta$ is the proportion used for the number of hub protein nodes in a network
\end{tablenotes}
\end{table*}

\begin{table*}
\centering
\caption{Simulation results under weak signal case using network-guided (NG) method, adaptive Lasso (aLasso), Lasso, elastic net (enet), ridge regression, and correlation-based penalized estimators (CBPE). The best results are highlighted in boldface.}
\label{sim.res2}
\begin{tabular}{@{}l cc l cccc @{}}
\toprule
Setting & $n$ & $p$ & Method & RMSE & CSL & F1 score & MCC \\
\midrule 
$n<p$ & 50 & 60 & NG ($\delta=0.06$)  & 0.26 (0.10) & \textbf{1.01} (0.02) & \textbf{0.88} (0.08) & \textbf{0.86} (0.09) \\
   &    && NG ($\delta=0.08$)  & 0.23 (0.10) & \textbf{1.01} (0.02) & 0.87 (0.07) & 0.85 (0.08)\\
   &    && NG ($\delta=0.10$)  & \textbf{0.20} (0.08) & \textbf{1.01} (0.01) & 0.86 (0.06) & 0.83 (0.06) \\
   &    && aLasso              & 0.49 (0.11) & 1.07 (0.04) & 0.80 (0.05) & 0.79 (0.06) \\
   &    && Lasso               & 0.34 (0.14) & 1.05 (0.03) & 0.62 (0.09) & 0.58 (0.11) \\
   &    && enet                & 0.46 (0.14) & 1.07 (0.04) & 0.53 (0.07) & 0.48 (0.09) \\
   &    && ridge               & 2.06 (0.11) & 3.43 (3.41) & 0.25 (0.00) & -- \\
   &    && CBPE                & 0.89 (0.12) & 0.97 (0.04) & 0.25 (0.00) & -- \\
\midrule
$n > p$ & 100 & 60 & NG ($\delta=0.06$)  & \textbf{0.08} (0.01) & 1.01 (0.00) & 0.96 (0.02) & 0.96 (0.03) \\
   &    && NG ($\delta=0.08$)  & \textbf{0.08} (0.01) & \textbf{1.00} (0.00) & 0.95 (0.01) & 0.94 (0.01) \\
   &    && NG ($\delta=0.10$)  & \textbf{0.08} (0.01) & \textbf{1.00} (0.00) & 0.91 (0.02) & 0.89 (0.02) \\
   &    && aLasso              & \textbf{0.08} (0.01) & 1.01 (0.00) & \textbf{1.00} (0.00) & \textbf{1.00} (0.00) \\
   &    && Lasso               & \textbf{0.08} (0.01) & 1.02 (0.00) & 0.78 (0.07) & 0.77 (0.07) \\
   &    && enet                & 0.09 (0.01) & 1.02 (0.00) & 0.64 (0.07) & 0.62 (0.07) \\
   &    && ridge               & 0.54 (0.05) & 1.10 (0.03) & 0.25 (0.00) & -- \\
   &    && CBPE                & 0.36 (0.04) & 0.96 (0.01) & 0.25 (0.00) & -- \\
\midrule
$n \ll p$ &100 & 300 & NG ($\delta=0.01$)  & 0.28 (0.08) & \textbf{1.01} (0.01) & \textbf{0.87} (0.07) & \textbf{0.87} (0.07) \\
   &    && NG ($\delta=0.02$)  & \textbf{0.19} (0.06) & \textbf{1.01} (0.01) & 0.81 (0.02) & 0.80 (0.03) \\
   &    && NG ($\delta=0.03$)  & 0.20 (0.07) & \textbf{1.01} (0.01) & 0.70 (0.02) & 0.71 (0.03) \\
   &    && aLasso              & 0.45 (0.06) & 1.05 (0.02) & 0.81 (0.03) & 0.82 (0.03) \\
   &    && Lasso               & 0.46 (0.05) & 1.06 (0.02) & 0.44 (0.09) & 0.47 (0.07) \\
   &    && enet                & 0.53 (0.07) & 1.08 (0.03) & 0.33 (0.06) & 0.39 (0.05) \\
   &    && ridge               & 2.14 (0.12) & 1.83 (0.60) & 0.06 (0.00) & -- \\
   &    && CBPE                & 2.48 (0.22) & 0.73 (0.07) & 0.06 (0.00) & -- \\
\bottomrule
\end{tabular}
\begin{tablenotes}\scriptsize
\item [] Abbreviations: RMSE, root-mean-squared error; CSL, calibration slope
\item [] $\delta$ is the proportion used for the number of hub protein nodes in a network
\end{tablenotes}
\end{table*}

\section{Real Data Application}

\subsection{Clinical Proteomic Tumor Analysis Consortium Data}
A pre-processed MS-based proteomics data of the National Cancer Institute (NCI) Clinical Proteomic Tumor Analysis Consortium (CPTAC) was downloaded from the Proteomic Data Commons (PDC; \url{https://pdc.cancer.gov/pdc/cptac-pancancer}), which is one of the largest public repositories of proteogenomic data. In this paper, the Estimation of STromal and Immune cells in MAlignant Tumor tissues using Expression data (ESTIMATE) \citep{ESTIMATE} score is the outcome of interest. The ESTIMATE score is a sum of the scores of immune and stromal cells, the two main non-tumor components in the tumor microenvironment. It has been used in a variety of cancer studies such as osteosarcoma \citep{ESTIMATE.cancer1}, head-and-neck squamous cell carcinoma (HNSCC) \citep{ESTIMATE.cancer2}, and lung cancer \citep{ESTIMATE.cancer3}. The higher the score, the lower the purity of the tumor. The clinical covariates that are included in the modeling with proteins are age, sex, body mass index (BMI), cancer staging, and size of tumor.

\subsection{Analysis}
We analyzed 337 gene-level proteins that are subsets of B cell-immune module from 108 patients with HNSCC. B cell-immune module was the rarest that accounts only in 3$\%$ of identifed modules depicted in a recent study \citep{CPTAC.petralia}. Of note, B cells have the ability to promote humoral immunity through the production of antobodies and its presence has been associated with responses to immunotherapy in cancer studies \citep{Bcells1, Bcells2}.

\begin{figure}[!ht]
\centering
\includegraphics[scale=0.3]{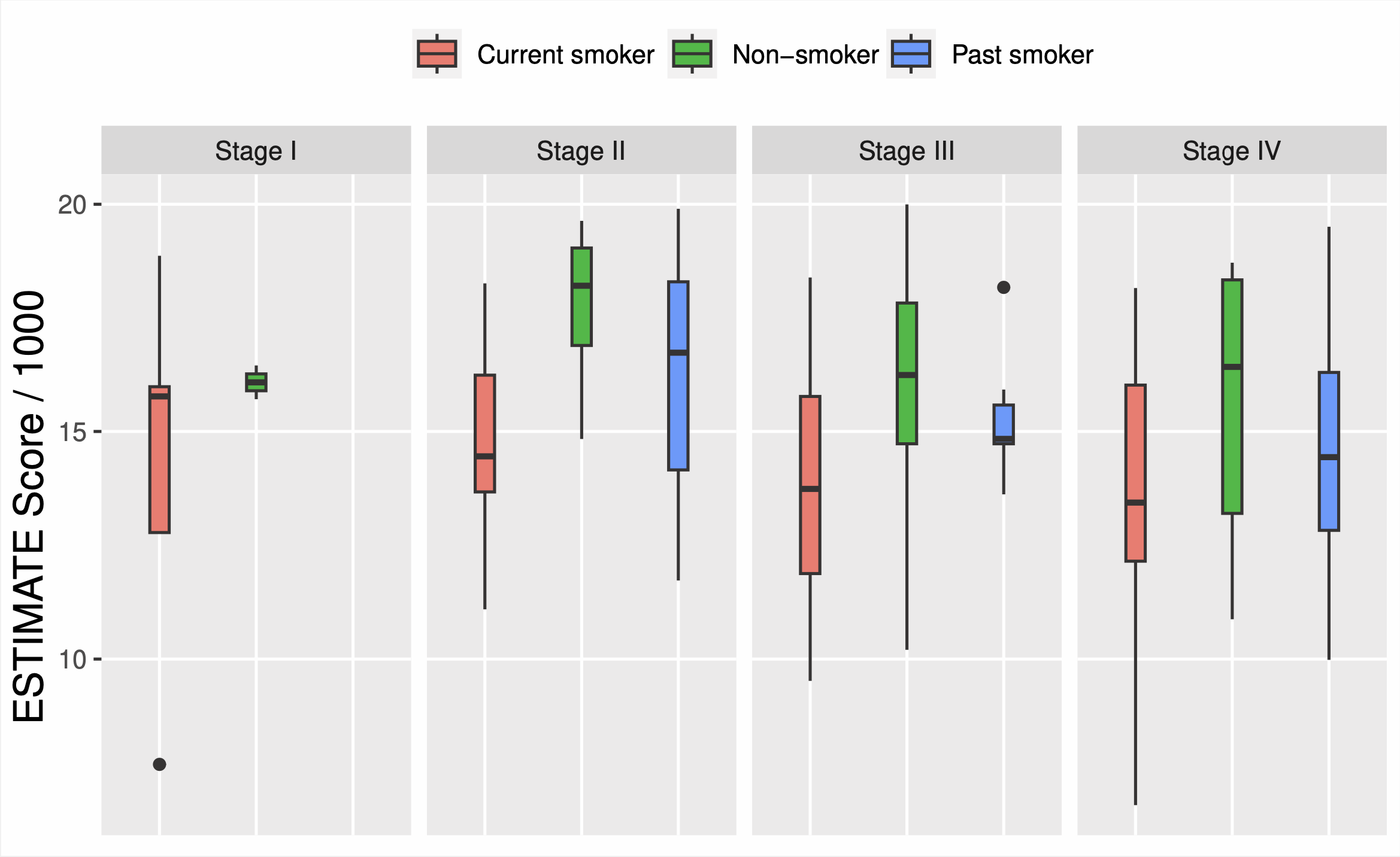}
\caption{Side-by-side boxplots to visualize the distributions of scaled ESTIMATE scores (scores divided by 1000) of CPTAC-HNSCC patients ($n =108$) by smoking status and cancer staging.}
\label{fig2}
\end{figure}

Table \ref{cptac.descriptive} summarizes characteristics of the full cohort sample, and we stratified these characteristics by smoking status. The study samples comprised older adults (median [IQR] = 62.0 [11.3]), predominantly male ($87\%$), and normal weights (24.0 [5.9]), according to the Centers for Disease Control and Prevention \citep{BMI}. 70.4$\%$ of patients were found in tumors staged III or IV with the median size of tumor was 3.2 cm (IQR=1.8). The ESTIMATE scores were significantly different across smoking statuses. In general, non-smokers had higher ESTIMATE scores in all cancer stages from early (stage I) to advanced disease (stage IV), when compared with current and past smokers. This is shown as side-by-side box plots in Figure \ref{fig2}.

Further, our proposed method was benchmarked on the CPTAC-HNSCC data against other popular methods that were evaluated in our simulation experiments above. When applying our method, three different values were considered for $\delta$, the proportion  of hub proteins in a network. We hypothesize that 3 ($\delta = 0.01)$, 6 ($\delta = 0.02)$, and 10 ($\delta = 0.03$) out of 337 gene-level proteins from B-cell immune module are defined as hubs in a network. As a whole, Table \ref{cptac.results} shows that our NG method had lower RMSE and better calibration slope (closer to 1) than that of the benchmark models. 

In addition to benchmark results, the HUGO Gene Nomenclature Committee (HGNC)-approved symbols of covariate-adjusted hub proteins are listed in Table \ref{cptac.list.proteomes}. HGNC-approved symbols are protein-coding gene annotations for each known human gene \citep{HGNC}. In this analysis, PABPC1, LGALS1, and GIMAP7 were found in common between three different values for $\delta$ parameter. By searching through integrative databases of human genes (GeneCards \citep{GeneCards}) and human diseases (MalaCards \citep{MalaCards}), PABPC1 is linked to viral diseases that are transmitted by mosquitoes such as rift valley fever and dengue virus. LGALS1 is related to corneal ulcer. Interestingly, a recent study \citep{GIMAP7} suggested that GIMAP7 has a potential as a prognostic biomarker and immune checkpoint gene for immunotherapy in pan-cancer. Additional hub proteins were identified when increasing the size of $\delta$. BLNK is associated with a rare genetic immunodeficiency disorder, called autosomal agammaglobulinemia \citep{BLNK}.

\begin{table*}
\centering
\caption{Patient characteristics of CPTAC-HNSCC patients.}
\label{cptac.descriptive}
\begin{tabular}{@{}lrrrrc@{}}
\toprule
Characteristics & \multicolumn{1}{c}{Overall}
& \multicolumn{1}{c}{Current smoker} & \multicolumn{1}{c}{Non-smoker}
& \multicolumn{1}{c}{Past smoker} & p-value$^{4}$ \\
& \multicolumn{1}{c}{N = 108} & \multicolumn{1}{c}{N = 61} & \multicolumn{1}{c}{N =20} & \multicolumn{1}{c}{N = 27} & \\
\midrule
Age in years$^{1}$ & 62.0 (11.3) & 62.0 (12.0) & 59.5 (11.3) & 64.0 (9.0)  & 0.2 \\
Sex$^{2}$    &  &  &  &   & 0.005 \\
 \hspace{3mm} Female & 14 (13.0\%) & 8 (13.1\%) & 6 (30.0\%) & 0 (0.0\%) & \\
 \hspace{3mm} Male & 94 (87.0\%) & 53 (86.9\%) & 14 (70.0\%) & 27 (100.0\%)  & \\
BMI$^{1}$  & 24.0 (5.9) & 24.0 (5.2) & 24.6 (6.2) & 24.0 (5.7)  & 0.005 \\
Cancer staging$^{2}$    &  &  &  &   & 0.7 \\
 \hspace{3mm} Stage I & 7 (6.5\%) & 5 (8.2\%) & 2 (10.0\%) &  0 (0.0\%) & \\
 \hspace{3mm} Stage II & 25 (23.1\%) & 14 (23.0\%) & 4 (20.0\%) &  7 (25.9\%) & \\
 \hspace{3mm} Stage III & 30 (27.8\%) & 16 (26.2\%) & 7 (35.0\%) & 7 (25.9\%) & \\
 \hspace{3mm} Stage IV & 46 (42.6\%) & 26 (42.6) & 7 (35.0\%) & 13 (48.1) & \\
Tumor size in cm$^{1}$     & 3.2 (1.8) & 3.0 (2.0) & 3.1 (1.4) &  4.0 (1.3) & 0.5 \\
ESTIMATE score$^{3}$     & 14.8 (4.0) & 13.9 (3.7) & 16.4 (3.5) &  15.2 (3.4) & 0.005 \\
\bottomrule
\end{tabular}
\begin{tablenotes}\scriptsize
\item[] Abbreviations: BMI, body mass index
\item [1] Median (IQR)
\item [2] $n (\%)$
\item [3] Scaled (score divided by 1000)
\item [4] Kruskal-Wallis test or Fisher's exact test as appropriate
\end{tablenotes}
\end{table*}

\begin{table*}
\centering
\caption{Analyses of CPTAC-HNSCC patients using network-guided (NG) method, adaptive Lasso (aLasso), Lasso, elastic net (enet), ridge regression, and correlation-based penalized estimators (CBPE). The best results are highlighted in boldface.}
\label{cptac.results}
\begin{tabular}{@{}lrrrrc@{}}
\toprule
 & \multicolumn{1}{c}{RMSE}
& \multicolumn{1}{c}{CSL} \\
\midrule 
NG ($\delta=0.01$)  & 1.95 &  \textbf{0.95} \\
NG ($\delta=0.02$)  & 1.78 &  \textbf{0.95} \\
NG ($\delta=0.03$) & \textbf{1.77} & 0.89  \\
aLasso  & 1.80 & 1.07 \\
Lasso  & 2.01 & 1.88 \\
Ridge  & 2.27 & 2.27  \\
enet & 1.90 & 1.51 \\
CBPE & 2.94 & 0.56 \\
\bottomrule
\end{tabular}
\begin{tablenotes}\scriptsize
\item [] Abbreviations: RMSE, root-mean-squared error; CSL, calibration slope
\item [] $\delta$ is the proportion used for the number of hub protein nodes in a network
\end{tablenotes}
\end{table*}

\begin{table*}
\centering
\caption{A complete list of significantly connected hub proteins that are adjusted by covariates. Results are shown by varying sizes of the proportion of hub proteins in a network ($\delta$). Listed proteins are mapped to gene symbols approved by the HUGO Gene Nomenclature Committee (HGNC). Proteins that are found in common between all three $\delta$ are boldfaced.\\}
\label{cptac.list.proteomes}
\begin{tabular}{@{}lrrrrc@{}}
\toprule
 & \multicolumn{1}{c}{HGNC Approved Symbols of Covariate-Adjusted Hub Proteins} \\
\midrule 
NG ($\delta=0.01$)  & \textbf{PABPC1}, \textbf{LGALS1}, \textbf{GIMAP7} \\
NG ($\delta=0.02$)  & \textbf{PABPC1}, \textbf{LGALS1}, \textbf{GIMAP7}, MEN1, RPLP1, HNRNPD \\
NG ($\delta=0.03$)  & \textbf{PABPC1}, \textbf{LGALS1}, \textbf{GIMAP7}, MEN1, RPLP1, HNRNPD, CASP10, \\ &  BLNK, SDC1, MUC4 \\
\bottomrule
\end{tabular}
\end{table*}

%However, we propose to utilize network-guided weights for the following reasons. First, the initial estimates to construct weights in adaptive Lasso may not be stable in an ultra-high dimensional setting where thousands of proteomes exist. Second, the network-guided weights takes into account conditional dependency of a pair of proteomes after controlling for all other nodes in a network, which represents an essential interacting structure between proteomes. 

\section{Discussion}
In this study, we have proposed a network-guided penalized regression model that retains hub proteins and clinical covariates, applying an adaptive Lasso penalty exclusively to non-hub proteins. This model screens out irrelevant non-hub proteins based on their predictive value, while maintaining key variables, including potential confounders for their clinical importance and hub proteins that are identified through network estimation.
Our hybrid method leverages network estimation and variable selection through partial penalization, representing a novel approach.
We have also shown that our network-guided estimators enjoy variable selection consistency and asymptotic normality.

Through a series of simulation studies and real data application, we have observed that the proposed network-guided approach demonstrates good overall performance measures. It is noteworthy that our method shines particularly in high-dimensional settings, where the number of proteins significantly exceeds the number of observations ($n \ll p$), which is a common scenario in most ``-omics'' disciplines.

Incorporating hub proteins is crucial, as they may serve as prognostic biomarkers across diverse diseases, including rare genetic disorders and immune checkpoints in cancer immunotherapy. Hub nodes can be identified based on various network properties. Here, we opted for degree centrality due to its intuitive nature and popularity in the literature. 
Future studies are warranted for the adoption of different network properties, such as betweenness and closeness centrality, to investigate how these impact results. 
It is also of interest to consider more flexible regression approaches which could handle repeatedly measured proteins data and/or complex relationships (e.g., non-linear or piece-wise) between covariates and the outcome of interest.

\printbibliography

\appendix

\section*{Appendix}
\subsection*{Proof of Theorem 1}

We first derive the joint asymptotic normality part. For any $\bm\alpha\in\mathbb{R}^{t}$, $\bm\beta\in\mathbb{R}^{q}$, denote $\bm{\theta}=(\bm\alpha^\T,\bm\beta^\T)^\T$,  $\hat{\bm{\theta}}_n=(\hat{\bm\alpha}_n^\T,\hat{\bm\beta}_n^\T)^\T$, and $\bm{\theta}_0=(\bm\alpha_0^\T,\bm\beta_0^\T)^\T$. Define $\bm{u} = \sqrt{n}(\bm{\theta} - \bm{\theta}_0)$ and $\hat{\bm{u}}_n = \sqrt{n} (\hat{\bm{\theta}}_n - \bm{\theta}_0)$. Then the objective function $L_n (\bm{\alpha}, \bm{\beta})$ can be re-written as
\begin{align*}
L_n (\bm{u}) 
& = \| Y - (\bm{U}, \bm{N})^\T \left( \bm{\theta}_0 + \frac{\bm{u}}{\sqrt{n}} \right) \|_2^2 + \lambda_n \sum_{s=t+1}^{t+q} \hat w_s \left| \theta_{0s} + \frac{u_s}{\sqrt{n}} \right|.
\end{align*}
It is then easy to verify that $\hat{\bm u}_n=\arg\min_{\bm u}L_n(\bm u)$. Note that $L_n (\bm{u}) - L_n (\bm{0}) = V_n (\bm{u})$ where
\begin{align*}
V_n (\bm{u}) = \bm{u}^\T n^{-1} \begin{pmatrix}
 \bm{U}^\T \bm{U} &  \bm{U}^\T \bm{N} \\ 
 \bm{N}^\T \bm{U} &  \bm{N}^\T \bm{N}
\end{pmatrix} \bm{u} - 2 & \frac{\epsilon^\T (\bm U, \bm N)}{\sqrt{n}}  \bm{u} \\
& + \frac{\lambda_n}{\sqrt{n}} \sum_{s=t+1}^{t+q} \hat w_s \sqrt{n} \left( \left| \theta_{0s} + \frac{u_s}{\sqrt{n}} \right| - \left|\theta_{0s}\right| \right).
\end{align*}
The first term in the above display converges to $\bm{u}^\T \bm{C u}$ for every $\bm u$. For the second term, by the Central Limit Theorem, we obtain $\epsilon^\T (\bm U, \bm N) /\sqrt{n}  \rightarrow_d \bm{W} = N(0, \sigma^2 \bm{C})$. The limiting behavior of the last term depends on whether $\theta_{0s}$ is active or not for $s=t+1, \ldots, t+q$, which is  equivalent to as whether $\beta_{0j}$ is active or not for $j=1, \ldots, q$. Note that if $\beta_{0j} \neq 0$, then $\tilde{\beta}_{nj} \rightarrow_p \beta_{0j}$ and $\hat w_j = |\tilde{\beta}_{nj} |^{-\nu} \rightarrow_p |\beta_{0j}|^{-\nu}$ by the Continuous Mapping Theorem. Also, $\sqrt{n} \left( \left| \beta_{0j} + \frac{u_j}{\sqrt{n}} \right| - \left|\beta_{0j}\right| \right) = u_j \textup{sgn} (\beta_{0j})$. Hence, by Slutsky's Theorem, $\frac{\lambda_n}{\sqrt{n}} \hat w_j \sqrt{n} \left( \left| \beta_{0j} + \frac{u_j}{\sqrt{n}} \right| - \left|\beta_{0j}\right| \right) \rightarrow_p 0$. If $\beta_{0j} = 0$, then $\sqrt{n} \left( \left| \beta_{0j} + \frac{u_j}{\sqrt{n}} \right| - \left|\beta_{0j}\right| \right) = |u_j|$ and $\frac{\lambda_n}{\sqrt{n}} \hat w_j = \frac{\lambda_n}{\sqrt{n}} n^{\nu/2} (|\sqrt{n} \tilde{\beta}_{nj}|)^{-\nu} \rightarrow \infty$ since $\sqrt{n} \tilde{\beta}_{nj} = O_p (1)$. Therefore, the last term converges in probability to $0$ if $\theta_{0s} \neq 0$, and it converges to $\infty$ if $\theta_{0s} = 0$. Now let $\mathcal{S} = \left \{1, 2, \ldots, t \right \} \cup \left \{s: \theta_{0s} \neq 0, \,s = t+1, \ldots, t+q \right \}$. %$\bm{u}_\mathcal{S}=\{u_s: s\in\mathcal{S}\}$, and $\bm{W}_\mathcal{S}$ is the part of $\bm W$ with indices in $\mathcal{S}$. 
Then by Slutsky's Theorem, we get $V_n (\bm{u}) \rightarrow_d V (\bm{u})$ for every $\bm{u}$, where
\[
V (\bm{u}) = \left\{\begin{matrix}
\begin{array}{ll}
\bm{u}_\mathcal{S}^\T \bm{C}_{\mathcal{S}} \bm{u}_\mathcal{S} - 2 \bm{u}^\T_\mathcal{S} \bm{W}_\mathcal{S} & \textup{if}\,\, u_s=0 \mbox{ for } s\notin \mathcal{S} \\ 
\infty & \textup{otherwise} 
 \end{array}
\end{matrix}\right.
\]
Note that $V (\bm{u})$ is convex and the minimum of $V (\bm{u})$ is uniquely achieved at $( \bm{C}_\mathcal{S}^{-1} \bm{W}_{\mathcal{S}},  \bm{0} )^\T$ where $\bm{C}_\mathcal{S}^{-1} \bm{W}_{\mathcal{S}} \in \mathbb{R}^{t+r}$ and $\bm{0} \in \mathbb{R}^{q-r}$. By the epi-convergence results of \cite{geyer} and \cite{knight}, we have 
\begin{align} \label{third}
\hat{\bm{u}}_{n\mathcal{S}} \rightarrow_d \bm{C}_\mathcal{S}^{-1} \bm{W}_{\mathcal{S}} \quad \textup{and} \quad \hat{\bm{u}}_{n\mathcal{S}^c} \rightarrow_d \bm{0}.
\end{align}
Therefore, $\hat{\bm{u}}_{n\mathcal{S}} = \sqrt{n} \binom{\hat{\bm{\alpha}}_n - \bm{\alpha}_0}{ \hat{\bm{\beta}}_{n\mathcal{J}} - \bm{\beta}_{0\mathcal{J}}} \rightarrow_d \bm{C}_\mathcal{S}^{-1} \bm{W}_{\mathcal{S}} = N (\bm{0}, \sigma^2 \bm{C}_\mathcal{S}^{-1})$, where $\bm{C}_{\mathcal{S}} \in \mathbb{R}^{(t+r) \times (t+r)}$ is the top-left block matrix (i.e., sub-matrix) of $\bm C \in \mathbb{R}^{(t+q) \times (t+q)}$.

Now we show the consistency part. Note that the asymptotic normality results imply that $\hat{\bm{\alpha}}_{n} \rightarrow_p \bm{\alpha}_{0}$ and $\hat{\beta}_{nj} \rightarrow_p \beta_{0j}$ for $\forall j \in \mathcal{J}$,  and hence $P(j \in \hat{\mathcal{J}}_n) \rightarrow 1$. Then it suffices to show that $\forall j' \notin \mathcal{J}$, $P(j' \in \hat{\mathcal{J}}_n) \rightarrow 0$. When $j' \in \hat{\mathcal{J}}_n$, we observe that $2 N_{j'}^\T (Y - \bm{U} \hat{\bm{\alpha}}_n - \bm{N} \hat{\bm{\beta}}_n) = \lambda_n \hat w_{j'} \textup{sgn} (\hat{\beta}_{nj'})$ by the Karush–Kuhn–Tucker (KKT) conditions. 
Note that $\lambda_n \hat w_{j'} \textup{sgn} (\hat{\beta}_{nj'}) / \sqrt{n} = \frac{\lambda_n}{\sqrt{n}} n^{\nu/2} \frac{1}{|\sqrt{n}\hat{\beta}_{nj'}|^{\nu}} \textup{sgn} (\hat{\beta}_{nj'}) \rightarrow_p \infty$, whereas
\begin{align*}
2 N_{j'}^\T (Y - \bm{U} \hat{\bm{\alpha}}_n -\bm{N} \hat{\bm{\beta}}_n )/\sqrt{n} & = 2 N_{j'}^\T \bm{U} \sqrt{n} (\bm{\alpha}_0 - \hat{\bm{\alpha}}_n )/n \\
 & \qquad + 2 N_{j'}^\T  \bm{N} \sqrt{n} (\bm{\beta}_0 - \hat{\bm{\beta}}_n ) /n + 2 N_{j'}^\T \epsilon / \sqrt{n}.
\end{align*}
By \eqref{third} and Slutsky's Theorem, we know that $2 N_{j'}^\T \bm{U} \sqrt{n} (\bm{\alpha}_0 - \hat{\bm{\alpha}}_n )/n$ and $2 N_{j'}^\T  \bm{N} \sqrt{n} (\bm{\beta}_0 - \hat{\bm{\beta}}_n ) /n$ converges in distribution to some normal distribution and $2 N_{j'}^\T \epsilon / \sqrt{n} \rightarrow_d N(\bm 0, 4 \|N_{j'}\|_2^2 \sigma^2)).$ Hence, $P(j' \in \hat{\mathcal{J}}_n) \leq P\left( 2 N_{j'}^\T (Y - \bm{U} \hat{\bm{\alpha}}_n -\bm{N} \hat{\bm{\beta}}_n) = \lambda_n \hat w_{j'} \textup{sgn} (\hat{\beta}_{nj'} ) \right) \rightarrow 0$. This proves the consistency part. %\hfill $\square$

%\addcontentsline{toc}{section}{Appendix}

\pagebreak

\pagebreak

\end{document}